\documentclass[twocolumn]{aastex62}

\usepackage{apjfonts}

\shorttitle{Nebula around SY Cancri}
\shortauthors{Bond et al.}

\newcommand{\Ha}{H$\alpha$}

\newcommand{\kms}{{\>\rm km\>s^{-1}}}

\def\nii{\ion{N}{2}}

\def\oiii{\ion{O}{3}}

\def\sii{\ion{S}{2}}

\def\Gaia{{\it Gaia}}

\newcommand{\TESS}{{\it TESS}}

\def\sycnc{SY~Cnc}

\begin{document}

\title{Discovery of a Bow-Shock Nebula around the Z~Cam-type Cataclysmic Variable SY Cancri}

\author[0000-0003-1377-7145]{Howard E. Bond}
\affil{Department of Astronomy \& Astrophysics, Penn State University, University Park, PA 16802, USA}
\affil{Space Telescope Science Institute, 
3700 San Martin Dr.,
Baltimore, MD 21218, USA}

\author[0009-0008-5193-4053]{Calvin Carter}
\affil{7215 Paldao Dr.,
Dallas, TX 75240, USA }

\author[0009-0000-4413-6434]{David F. Elmore}
\affil{National Solar Observatory, 3665 Discovery Dr., Boulder, CO 80303, USA}

\author[0009-0005-3715-4374]{Peter Goodhew}
\affil{Deep Space Imaging Network, 108 Sutton Court Rd., London, W4 3EQ, UK}

\author[0000-0002-9018-9606]{Dana Patchick}
\affil{Deep Sky Hunters Consortium, 1942 Butler Ave. Los Angeles, CA 90025, USA }

\author[0009-0009-3986-4336]{Jonathan Talbot}
\affil{Stark Bayou Observatory, 1013 Conely Cir., Ocean Springs, MS 39564, USA}

\correspondingauthor{Howard E. Bond}
\email{heb11@psu.edu}

\begin{abstract}

We report the serendipitous discovery of a bow-shock nebula around the cataclysmic variable (CV) SY~Cancri. In addition, SY~Cnc lies near the edge of a faint \Ha-emitting nebula with a diameter of about $15'$. The orientation of the bow shock is consistent with the direction of SY~Cnc's proper motion. Nebulae are extremely rare around CVs, apart from those known to have undergone classical-nova (CN) outbursts; bow shocks and off-center nebulae are even more unusual. Nevertheless, the properties of SY~Cnc and its nebulosity are strikingly similar to those of V341~Ara, another CV that is also associated with a bow shock and is likewise off-center with respect to its faint \Ha\ nebula. Both stars are binaries with optically thick accretion disks, belonging to the classes of Z~Cam CVs or nova-like variables. We discuss three scenarios to explain the properties of the nebulae. They may have resulted from chance encounters with interstellar gas clouds, with the stars leaving in their wakes material that is recombining after being photoionized by UV radiation from the CVs. Alternatively, the large nebulae could be ejecta from unobserved CN outbursts in the recent past, which have been decelerated through collisions with the interstellar medium (ISM), while the stars continue to snowplow through the gas. Or the faint \Ha\ nebulae may be ambient ISM that was shock-ionized by a CN outburst in the past and is now recombining.

\null\vskip 0.2in

\end{abstract}



\section{Introduction: Nebulae and Bow Shocks Around Cataclysmic Variable Stars \label{sec:intro} }

Cataclysmic variables (CVs) are short-period binary systems in which a Roche-lobe--filling star transfers mass to a white-dwarf (WD) companion. In the rare CV subclass of magnetic AM~Herculis systems, the transferred material accretes directly onto the WD\null. However, in the majority of CVs the material resides in an accretion disk around the WD before eventually falling onto it. The main subclasses of non-magnetic CVs include the classical novae (CNe, in which the accreted hydrogen ignites nuclear fusion on the surface of the WD); dwarf novae (DNe, in which the accretion disk is usually optically thin, but becomes optically thick during occasional DN outbursts); and nova-like variables (NLVs, in which the transfer rate is so high that the accretion disk remains optically thick most or all of the time). An important subset of the DNe is the Z~Cam variables \citep[ZCVs; see][]{Simonson2014}, which exhibit DN outbursts, but can occasionally remain in ``standstills'' at an intermediate brightness level for a few days up to more than 1000~days. For comprehensive reviews of CVs, see the monographs and papers by \citet{Warner1995}, \citet{Szkody2012}, and \citet{Sion2023}. 

Following a CN outburst, the central binary is often surrounded by an expanding nebula, which can be visible for several decades or even more than a century \citep[e.g.,][and references therein]{Tappert2020, Santamaria2020}. Eventually the nebula fades away as the nova ejecta dissipate into the interstellar medium (ISM)\null. Several NLVs and DNe that are not known to have undergone CN eruptions have been found to be surrounded by very faint nebulae, often with a filamentary structure. These include the ZCVs Z~Cam itself \citep{Shara2007, Shara2024} and AT~Cnc \citep{Shara2012, Shara2017}, and the NLVs IPHASX J210204.7+471015 \citep{Guerrero2018}, and V1315~Aql \citep{Sahman2018}. The cited authors have plausibly argued that these nebulae were ejected during unobserved CN outbursts of the central stars several centuries ago. 

Except for these shells surrounding known and suspected old novae, emission nebulae around CVs are extremely and somewhat puzzlingly rare \citep[see][]{Tappert2020}. For example, \citet{Sahman2015}, \citet{Schmidtobreick2015}, \citet{Pagnotta2016}, and \citet{Sahman2022} searched for \Ha-emitting nebulae around a combined total of $\sim$150 CVs, with an emphasis on NLVs and ZCVs. They found only one convincing new example, a nebula with a radius of $\sim$$2\farcm5$ centered on the NLV V1315~Aql, which is mentioned in the previous paragraph. 

In a very few cases, CVs have been found to be associated with bow-shock nebulae. The shock is formed when the CV encounters an interstellar cloud at a relatively high velocity, and a fast wind arising from its accretion disk collides with the surrounding ISM\null. Probably the best-known such object is the faint parabolic bow-shock nebula EGB~4, discovered by \citet{Ellis1984}. These authors found that a blue star lies near the apex of the parabola. It proved to be a previously unknown NLV, subsequently designated BZ~Cam. Deep emission-line images of EGB~4 have been presented in papers by \citet{Hollis1992}, \citet{Greiner2001}, and \citet[][hereafter BM18]{BondV341Ara2018}, and online by coauthor P.G.\footnote{\url{https://www.astrobin.com/5m82dg/C}} An extraordinary extremely deep wide-angle image of EGB~4 has been obtained by amateur Niko Giesriegler,\footnote{\url{https://www.astrobin.com/7ct9sv}} revealing a long, faint comet-like tail extending at least $14'$ north of BZ~Cam (a projected $\sim$1.5~pc at the distance of the star), in a direction opposite to the proper motion of the star.

A remarkably similar wind-driven bow-shock nebula, associated with the variable star V341~Ara, was discovered by \citet{Frew2008}. V341~Ara had been considered, since its discovery in the early 20th century by Henrietta Leavitt, to be a Cepheid variable. However, Frew showed that it is actually a NLV, based on spectroscopy, photometry, and its identification with an X-ray source. Deep images of the nebula have been presented by BM18 and \citet[][hereafter CS21]{CastroSegura2021}. V341~Ara and its bow shock lie at the edge of an extended faint \Ha-emitting nebula, designated Fr~2-11, which has a diameter of about 0.3~pc. 

A few other bow shocks around CVs have been discovered in recent years. The faint nebula around the NLV IPHASX J210204.7+471015, mentioned above, has bow-shock features, interpreted by \citet{Guerrero2018} as due to an interaction between ejecta from an unobserved CN eruption and the ISM\null. A very faint bow shock around a newly discovered DN, V1838~Aql, was detected by \citet{Hernandez2019}. This CV has a very high transverse velocity of $\sim$$123\,\kms$. Another possible bow shock around the newly discovered NLV ASASSN-V J205457.73+515731.9 was pointed out by \citet{BondASASSN2020}. Like V341~Ara, this CV lies at the edge of an extended, faint \Ha\ nebula.

In this paper, we report the accidental discovery by amateur astronomers of a bow shock and faint \Ha\ nebula around a well-known and bright Z~Cam star, SY Cancri.

\section{A Serendipitous Discovery \label{sec:discovery} }

Author D.P. has successfully conducted searches for faint planetary nebulae (PNe) for many years, as a member of the Deep Sky Hunters Consortium (see, for example, \citealt{Jacoby2010}, \citealt{Kronberger2016}, and \citealt{Ritter2023}). His explorations are based on available online sky surveys. In the course of examining wide-angle images at high Galactic latitudes from The Southern H-Alpha Sky Survey Atlas\footnote{SHASSA images can be downloaded from \url{https://skyview.gsfc.nasa.gov/current/cgi/query.pl}} \citep[SHASSA;][]{Gaustad2001}, he noticed a field overlain with extensive faint \Ha\ emission in the constellation Cancer. This suggested the presence of a relatively nearby and ancient PN, worth further investigation. 

This candidate PN was pointed out to collaborators P.G. and D.F.E. for followup imaging. On 2022 January 27-28, D.F.E. used a 55-mm aperture $f/3.6$ refractor\footnote{\url{https://kenkoglobal.com/lp/borg/page28357607.html}} with a ZWO\footnote{\url{https://www.zwoastro.com/}} ASI2400MC color camera and a dual \Ha\ and [\oiii] 5007~\AA\ filter\footnote{\url{https://idas.uno/space/en/IDAS/nbz.htm}} to obtain 52 exposures of 600~s each. D.F.E.'s telescope is located at the Dark Sky New Mexico\footnote{\url{https://darkskynewmexico.com/}} site. These exposures confirmed the extended \Ha\ emission across the field---but upon examining the images, D.P. noticed that in the extreme northwest corner of the wide-angle image there was an uncataloged faint and roughly circular nebula\footnote{This object lies just outside the northernmost coverage of the area by SHASSA.} with a diameter of about $15'$. It appeared to coincide with a known variable star, SY~Cancri. D.P. contacted D.F.E. and urged him to follow up on this discovery.

D.F.E. then imaged the SY~Cnc nebula on 2022 January 30-31, this time using his 100-mm aperture $f/3.8$ refractor\footnote{\url{https://global.vixen.co.jp/en/product/26145_1/}} and a ZWO ASI6200MM monochrome camera with narrow-band emission-line filters. Frames of 600~s each were obtained in \Ha, [\oiii], and [\sii] 6716--6731~\AA\ Chroma filters\footnote{\url{https://www.chroma.com/}} (21, 17, and 15 exposures, respectively). These images confirmed the reality of the nebula, and indicated the likely presence of a bow shock. The frames were posted to an amateur website\footnote{\url{https://planetarynebulae.net/EN/page_np.php?id=1044}} devoted to searches for faint PNe. On this basis the nebula was designated PaEl\,1, and it was added as a candidate PN to the online Hong-Kong/AAO/Strasbourg/H$\alpha$ Planetary Nebulae (HASH) database\footnote{\url{http://hashpn.space/}} \citep{Parker2016, Bojicic2017}. 
It was the inclusion of PaEl~1 in HASH that caught the attention of H.E.B., who is conducting a spectroscopic survey of central stars of newly discovered faint PNe with the Hobby-Eberly Telescope \citep[see][and subsequent papers in that series]{BondPaperI2023}. In this case, however, the star was already a well-known object, and of a type not normally seen as nuclei of PNe.

\section{Deep Imaging of P\MakeLowercase{a}E\MakeLowercase{l} 1 \label{sec:deep_imaging} }

\begin{figure*}
\centering
\includegraphics[width=5in]{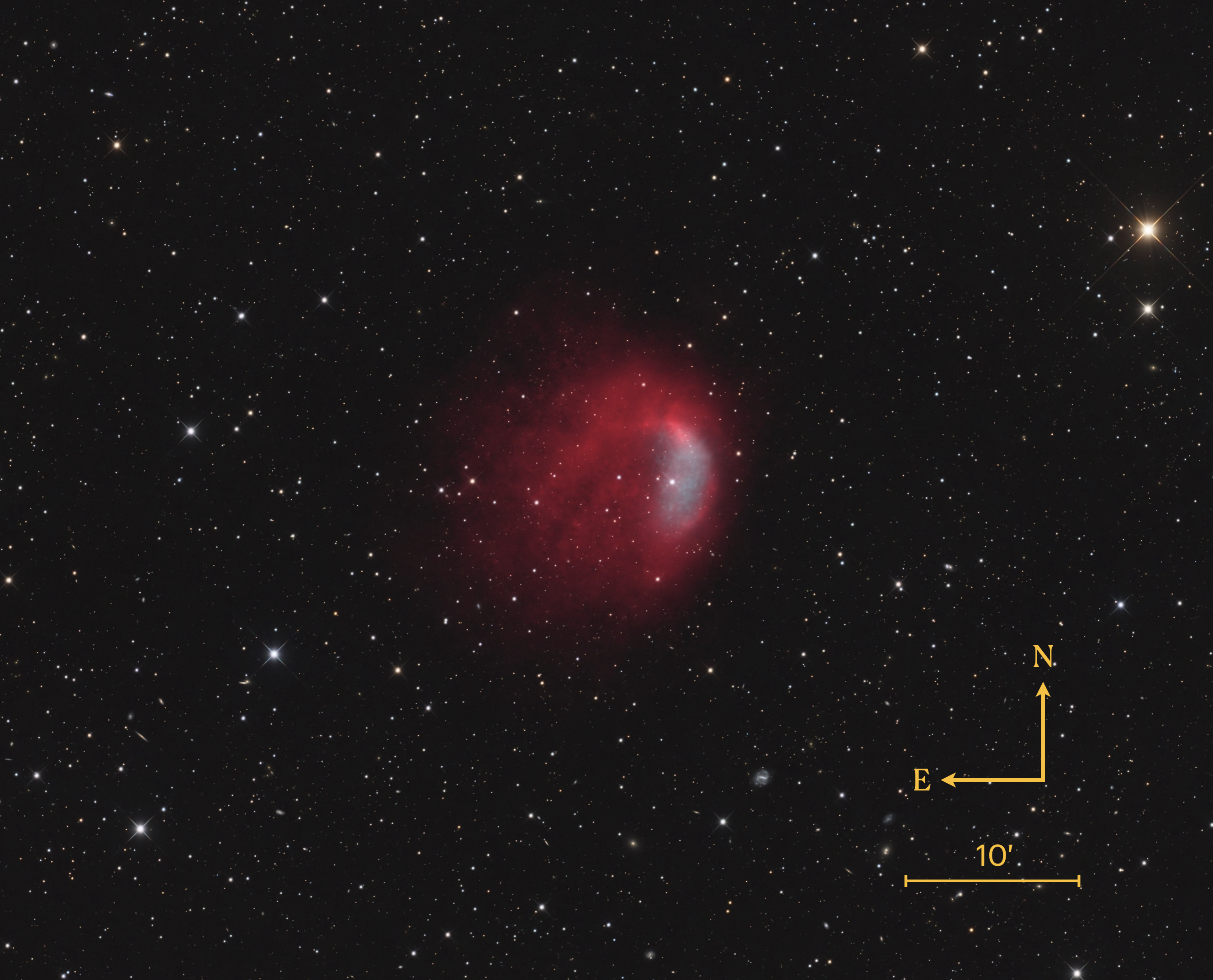}
\caption{
Deep image of PaEl\,1, from a total of 16.76 hours of exposure with a 350-mm $f/3$ telescope (see Section~\ref{sec:deep_imaging} for details). Height of frame is $54'$. RGB filters are used for the stellar background, and narrow-band frames in \Ha\ and [\oiii] $\lambda$5007 are mapped to red and blue-green, respectively. Scale and orientation are indicated in the picture. The bright star just east of the center of the [\oiii] emission is the cataclysmic variable star SY~Cnc.
\label{fig:color_image}
}
\end{figure*}

In order to investigate PaEl\,1 in more detail, coauthors J.T. and C.C. obtained deep high-resolution images with a PlaneWave DeltaRho 350-mm aperture $f/3$ telescope,\footnote{\url{https://planewave.com/products/deltarho-350-ota}} located at the Dark Sky Observatory\footnote{\url{http://darkskyobservatory.com/index.html}} in west Texas, USA\null. The imager is a ZWO ASI461MM Pro CMOS camera. Exposures were accumulated between 2024 April~22 and June~3, and totaled 50~min ($50\times60$~s) each in RGB filters, and 10.27 ($77\times480$~s) and 4.0~hr ($30\times480$~s) in Chroma \Ha\ and [\oiii] filters, respectively. The resulting image, with red mapped to \Ha\ and blue-green to [\oiii], is shown in Figure~\ref{fig:color_image}. Additional technical details of the exposures and processing for this image are available online.\footnote{At \url{https://www.starscapeimaging.com/PaEl 1/PaEl1.html} and \url{https://www.astrobin.com/0uqr8r/}}

Two further images of this extremely faint nebula have been posted online recently, one by Sven Eklund,\footnote{\url{ https://www.astrobin.com/lr4g62/}} and the other by P.G\null. The latter's image\footnote{\url{https://www.imagingdeepspace.com/pael-1.html}} reaches a similar depth as in Figure~\ref{fig:color_image}. In retrospect, the nebula is actually marginally visible in photographic images from the Space Telescope Science Institute Digitized Sky Survey,\footnote{\url{https://archive.stsci.edu/cgi-bin/dss_form}} but it had escaped notice until now.

SY~Cnc is prominent in Figure~\ref{fig:color_image} as the bright star on the eastern side of an area of bright [\oiii] emission lying at the western edge of the \Ha\ emission of PaEl\,1. Table~\ref{tab:DR3data} gives the star's equatorial and Galactic coordinates, absolute parallax and proper-motion components, and mean apparent magnitude and color, all taken from \Gaia\/ Data Release~3\footnote{\url{https://vizier.cds.unistra.fr/viz-bin/VizieR-3?-source=I/355/gaiadr3}} (DR3; \citealt{Gaia2016, Gaia2023}). The \Gaia\/ parallax implies a distance of $\sim$405~pc.\footnote{A Bayesian analysis of {\it Gaia\/} EDR3 data by \citet{BailerJones2021} yields a distance of $401.2^{+5.0}_{-4.3}$~pc.} At this distance, the total proper motion of $29.84\pm0.03\rm\,mas\,yr^{-1}$ implies a relatively high transverse space motion of $56.8\pm0.6\,\kms$. The center-of-mass radial velocity of the star is only $-0.2\pm2.5\,\kms$ \citep{Casares2009}, so its motion is almost precisely in the plane of the sky, as seen from our location.


\begin{deluxetable}{lc}
\tablecaption{\Gaia\/ DR3 Data for \hbox{SY Cancri} \label{tab:DR3data} }
\tablehead{
\colhead{Parameter}
&\colhead{Value}
}
\decimals
\startdata
RA (J2000) & 09 01 03.320 \\
Dec (J2000) & +17 53 56.03 \\
$l$ [deg] &  210.01 \\
$b$  [deg] &  +36.44 \\
Parallax [mas] & $2.461\pm0.025$ \\
$\mu_\alpha$ [mas\,yr$^{-1}$] & $-29.138\pm0.028$ \\
$\mu_\delta$ [mas\,yr$^{-1}$] & $-6.423\pm0.021$ \\
$G$ [mag] &  12.69 \\
$G_{\rm BP}-G_{\rm RP}$ [mag] & $0.60$ \\
\enddata
\end{deluxetable}

\begin{figure}
\centering
\includegraphics[width=0.47\textwidth]{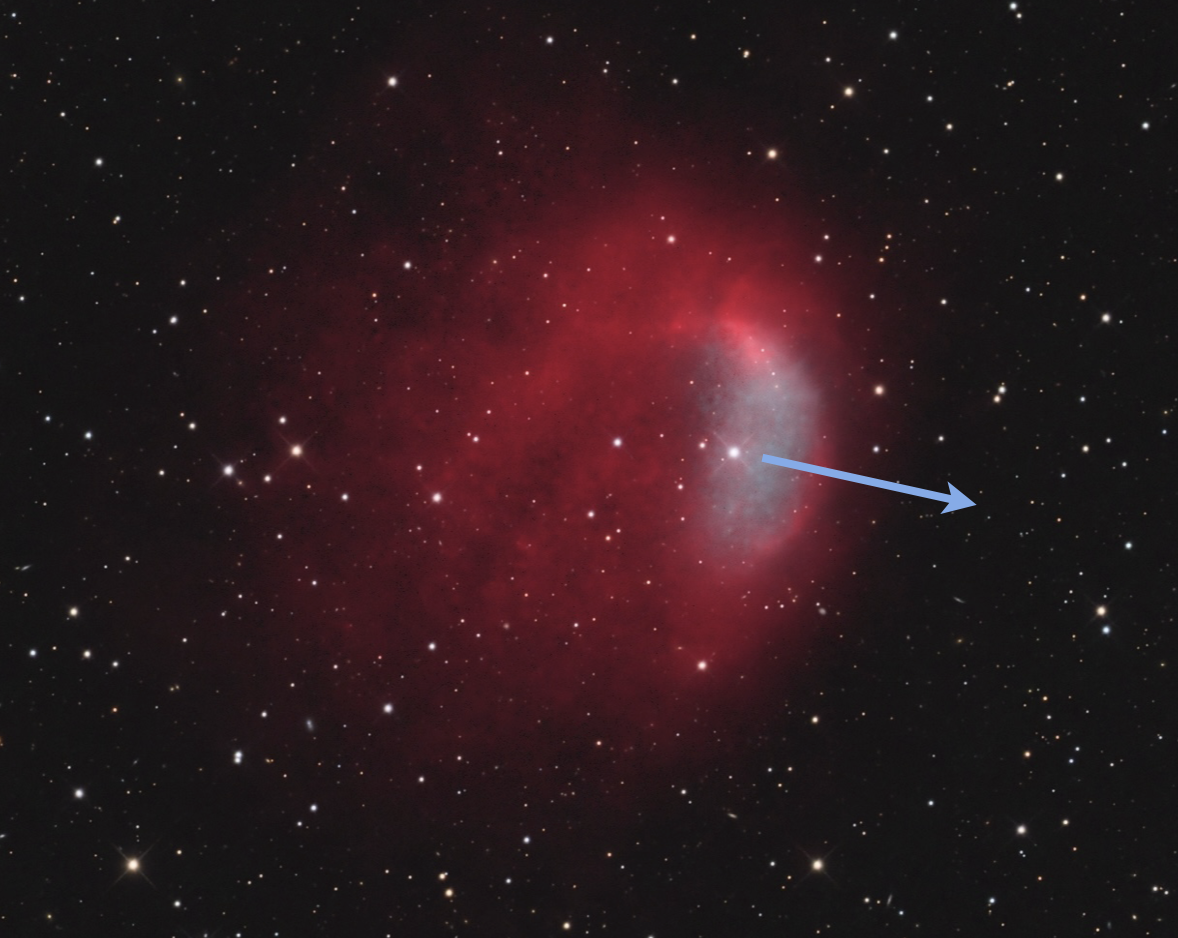}
\caption{
Close-up of PaEl~1. The arrow is in the direction of the proper motion of SY~Cnc.
\label{fig:zoom_in}
}
\end{figure}

Figure~\ref{fig:zoom_in} zooms in on the image of PaEl\,1 from Figure~\ref{fig:color_image}. Here we clearly see the morphology of a bow shock, in the form of a bright parabolic rim to the west of the star, which is prominent in the light of [\oiii], but also seen in \Ha. This is the signature of a fast wind from the star, colliding with the ISM as the star passes through it at a supersonic velocity. The blue arrow in Figure~\ref{fig:zoom_in} is in the direction of the proper motion of SY~Cnc, at position angle $257\fdg6$; it shows that the high-speed motion of the star is in a direction consistent with the interpretation as a bow shock. 

In the light of \Ha\ the surrounding nebula is large and roughly circular, but with a fainter extension that fades away to the east. Its angular diameter is roughly $15'$, about 1.8~pc at the distance of the star, but this is very approximate because the edges of the \Ha\ nebula are ill-defined. SY~Cnc lies very conspicuously off-center to the west with respect to the \Ha\ nebula, in the direction of the star's motion.\footnote{We are referring here to an ``\Ha\ nebula,'' but we note that the Chroma \Ha\ filter also transmits the neighboring [\nii] emission lines at 6548 and 6583~\AA, and they may be contributing significantly.} 


\section{The Z C\MakeLowercase{am} Variable Star SY~C\MakeLowercase{ancri} \label{SY_Cnc} }


The variability of \sycnc\ was discovered by Lydia Ceraski, and announced by \citet{Blazko1929}. She reported that the object varied between magnitudes 9.5 and 12.5 on 14 photographic plates obtained between 1912 and 1928. Magnitudes as bright as 9.5 seem surprising in view of modern data discussed below; however, Ceraski noted that the variable is identical with BD~+18$^\circ$2101, cataloged visually in the 19th century also at magnitude 9.5. In a followup study, \citet{Prager1931} reported 31 measurements, ranging from magnitude 9.9 to fainter than 12. Thus it seems possible that the star has historically become brighter than 10th magnitude. However, the Light Curve Generator\footnote{\url{https://www.aavso.org/LCGv2/}} of the American Association of Variable Star Observers (AAVSO) shows that SY~Cnc has varied between visual magnitudes of about 10.5 to 14 from 1956 to the present.\footnote{See also the AAVSO light curves of SY~Cnc plotted by \citet[][their Figure~3]{LandoltClem2018} and \citet[][his Figure~11]{Oshima2022}.}

The first spectroscopic study of SY~Cnc was carried out by \citet{Herbig1950}, who noted the similarity of its spectrum to that of Z~Camelopardalis---which is now recognized as the prototype of the ZCVs, as described in Section~\ref{sec:intro}. Based on its AAVSO light curve, and other information, SY~Cnc is listed as one of only 19 ``{\it bona fide}'' ZCVs by \citet[][their Table~1]{Simonson2014}; in fact, with a maximum brightness of 10.5~mag, it would be the third-brightest known ZCV\null. 

\begin{figure}[h]
\centering
\includegraphics[width=0.47\textwidth]{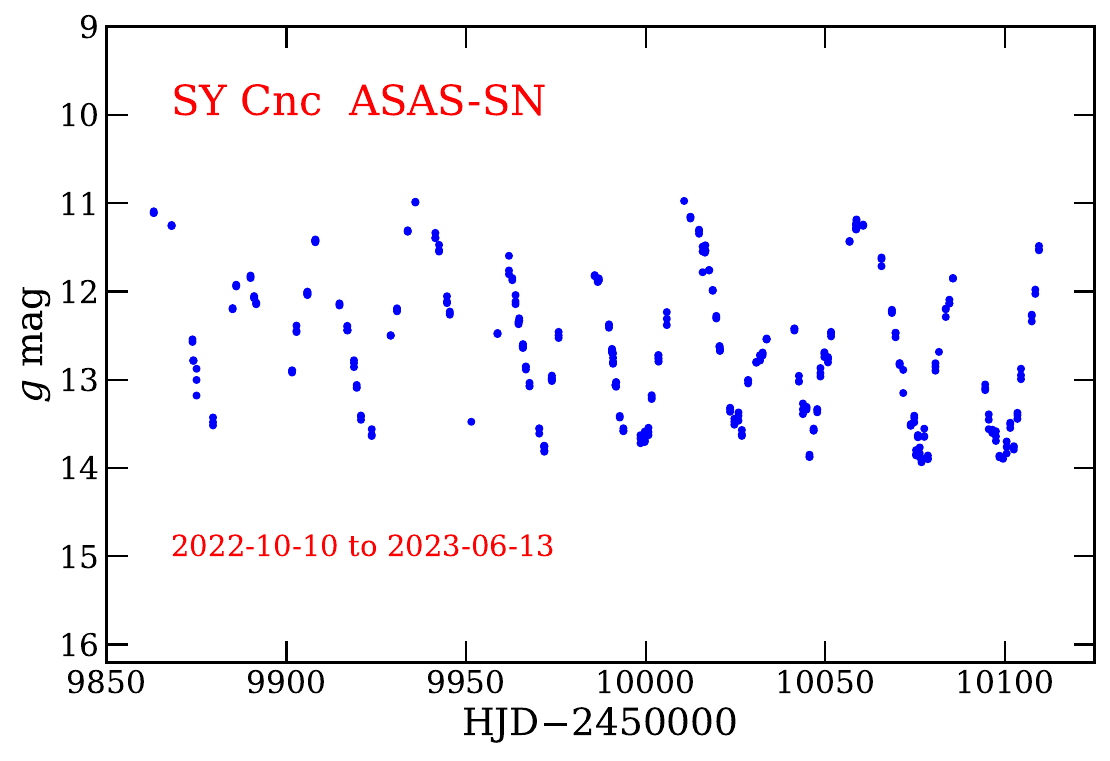}
\caption{
ASAS-SN $g$-band light curve of SY~Cnc over an eight-month interval in 2022-2023. The star shows a quasi-periodic sawtooth-shaped light curve, with a mean interval between outbursts of 25.3~days.
\label{fig:asas-sn}
}
\end{figure}

There is substantial astrophysical information available in the literature about SY~Cnc, and we only summarize here. An historical overview is given by \citet{LandoltClem2018} in their Section~1. ZCVs typically have regularly spaced outbursts---apart from the times when they enter into standstills---giving their light curves a ``sawtooth'' appearance \citep[see, e.g.,][]{Shafter2005}. SY~Cnc conforms to this behavior. As an example, we show in Figure~\ref{fig:asas-sn} its $g$-band light curve from the All-Sky Automated Survey for Supernovae\footnote{\url{https://asas-sn.osu.edu/}} (ASAS-SN; \citealt{Shappee2014}, \citealt{Kochanek2017}) for an eight-month interval during the 2022-2023 season. Here the outbursts are spaced at a mean interval of 25.3~days. Examining the ASAS-SN seasons between 2017-2018 and 2023-2024, we find that the mean outburst intervals have ranged from 25.3 to 29.4~days. Similar results were obtained by \citet{Shafter2005} in their investigation of AAVSO data covering about a decade, showing a median outburst interval of 26~days, and by \citet{Oshima2022}, who found a mean interval of 26.33~days from AAVSO data covering 1995 to 2022. The complete AAVSO light curve from 1956 to the present shows that this behavior has been remarkably consistent for almost seven decades, apart from a standstill observed in 1982 \citep[][their Figure~10]{Simonson2014}.

The orbital period of SY~Cnc was found to be $\sim$0.38~days by \citet{Shafter1983}, which has been refined to $0.3823753 \pm 0.0000003$~days (9.18~hr) by \citet{Casares2009}. This is one of the longest known periods for a ZCV\null. \citet{Casares2009} detected the donor star in their spectra of the system, showing it to be a late G-type star, which must be an evolved star in order for it to fill its Roche lobe at this long a period. The long orbital period also accounts for the relatively long outburst period, given the empirical correlation between orbital and outburst periods for ZCVs found by \citet{Shafter2005}.

\begin{figure}[h]
\centering
\includegraphics[width=0.47\textwidth]{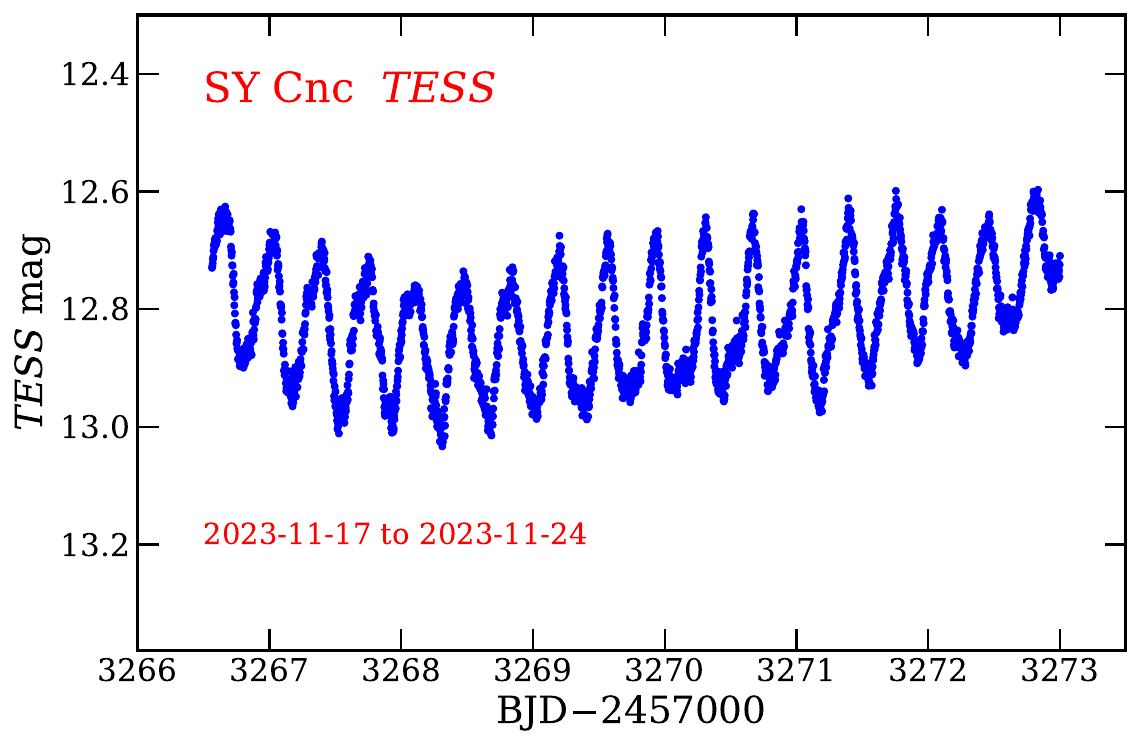}
\caption{
A representative \TESS\/ light curve for SY~Cnc, covering a 6.4-day interval at a 200~s cadence. Variations are seen with a period of 8.72~hr, about 5\% shorter than the orbital period of 9.18~hr. This phenomenon is called ``negative superhumps''; see text for discussion. 
\label{fig:tess}
}
\end{figure}

We examined high-precision aperture photometry of SY~Cnc obtained by the {\it Transiting Exoplanet Survey Satellite\/} (\TESS) mission, using the online {\tt TESSExtractor} tool\footnote{\url{https://www.tessextractor.app}} \citep{Serna2021}. Figure~\ref{fig:tess} shows a sample of the \TESS\/ data, in this case excerpted from the Sector~72 run, having a cadence of 200~s. Measurements for an interval of about 6.4~days are plotted, giving a representative view of the star's photometric behavior. Here we see another ``sawtooth'' light curve, with a peak-to-peak amplitude of about 0.3~mag, but with a much shorter period than the outburst period. The period for these variations is 8.72~hr, which is similar to the orbital period of 9.18~hr, but {\it less than\/} it by about 5\%.
Such oscillations at periods a few percent shorter than the orbital period are called ``negative superhumps,'' which are interpreted as arising from a beat between the orbital period and a retrograde precession of a tilted accretion disk. There are extensive discussions of this phenomenon in the CV literature; see, for example, \citet{Sun2023, Sun2024} and references therein.

\section{Discussion \label{discussion} }

We have discovered serendipitously that the variable star SY~Cnc is surrounded by a faint bow-shock nebula. Moreover, it lies near the edge of a larger, even fainter, and roughly circular \Ha\ nebula, designated PaEl\,1. In the previous section we reviewed the properties of the star. It is a member of the class of Z~Cam CVs, with an orbital period of 9.18~hr (one of the longest periods known among the ZCVs), showing quasi-regularly spaced outbursts at an interval typically of about 26~days, along with negative superhumps at a period about 5\% shorter than the orbital period.

Many of these properties are quite strikingly similar to those of another object, the NLV V341~Ara. As discussed in the studies of BM18 and CS21, referenced in Section~\ref{sec:intro}, this CV is likewise accompanied by a bow shock, and it similarly lies at the edge of a larger, faint \Ha-emitting nebula, Fr~2-11. The stars' mean absolute magnitudes, based on the parallaxes and magnitudes in \Gaia\/ DR3, are nearly identical: $M_G=+4.6$ for SY~Cnc, and +4.8 for V341~Ara.\footnote{These absolute magnitudes are uncorrected for interstellar reddening, but it is small and nearly the same for both stars, about $E(B-V)=0.03$, according to the {\tt Stilism} tool at \url{https://stilism.obspm.fr}. BZ~Cam is only slightly fainter, at a mean $M_G=+5.1$. The absolute magnitude of ASASSN-V J205457.73+515731.9, a possible third member of this class of off-center CVs in faint nebulae, mentioned in Section~\ref{sec:intro}, is also a similar $M_G=+4.7$.} These are much brighter than DNe in quiescence, and are indicative of optically thick accretion disks. However, the orbital period of V341~Ara, 3.65~hr, is quite a bit shorter than that of SY~Cnc; and it has quasi-regular outbursts whose timescales of $\sim$10--16~days are shorter than those of SY~Cnc. This is consistent with the correlation between orbital and outburst periods discussed by \citet{Shafter2005}. Like SY~Cnc, \TESS\/ photometry of V341~Ara reveals negative superhumps, with a period about 1\% shorter than its orbital period (CS21).

BM18 and CS21 considered two scenarios to explain why V341~Ara exhibits a bow shock that is embedded close to the edge of a fainter and larger nebula---and these scenarios should apply as well to the remarkably similar structures we have discovered around SY~Cnc. They are as follows:

1.~In the first scenario, V341~Ara and SY~Cnc (and BZ~Cam as well) are CVs undergoing a chance high-velocity encounter between the binary system and an interstellar gas cloud. All three of these variables are ZCVs or NLVs; they have high mass-transfer rates from donor star to WD, producing optically thick accretion disks, and a significant rate of mass ejection from the disk into space as a high-speed wind. The collision between the wind and the essentially stationary gas produces the conspicuous bow shocks seem in all three objects. As the CV passes through the surrounding large cloud, its ultraviolet (UV) radiation photoionizes the gas, producing the large and diffuse \Ha\ clouds lying in the ``wake'' behind the variable star's path.

2.~In an alternative scenario, the large \Ha\ nebulae are not random interstellar clouds, but are material that was actually ejected from the CVs during CN outbursts in the (recent) past. This scenario requires that the nova ejecta---which would initially have shared the fast center-of-mass motion of the star---must have been slowed down substantially through interactions with the ISM\null. This is necessary so that the wind from the star, whose space motion has suffered no deceleration, is able to produce the observed bow shock as its wind ``snowplows'' through the star's own ejected material.

In the first scenario, in which UV radiation from SY~Cnc photoionizes an interstellar cloud, the extended nebula will be recombining on the eastern edge---where in fact Figures~\ref{fig:color_image} and \ref{fig:zoom_in} do show the nebula gradually fading away with distance from the star, as opposed to the relatively sharper edges on the northern, western, and southern sides. This interpretation should be testable with spectroscopic observations of the nebula; for example, we would expect [\nii] to be strong relative to \Ha\ where the recombination is occurring on the eastern edge. However, these observations would be difficult since the nebulosity here is extremely faint.

If we adopt the second scenario, it is possible to make a rough estimate of how long ago the nova eruptions occurred, since we know the proper motion of the star and its angular separation from the center of the \Ha\ nebula. BM18 and CS21 did this for V341~Ara, under the assumption that the ejected \Ha\ nebula has been decelerated to essentially zero motion. In this case, the nova explosion for V341~Ara occurred about 800--1000~yr ago. A similar calculation for SY Cnc yields an outburst very approximately 7,800~yr ago.\footnote{CS21 discussed a very tentative identification of V341~Ara with a ``guest star'' recorded by Chinese astronomers in the year 1240, but this appears dubious because the star lies at a declination of $-63^\circ$. In contrast, SY~Cnc is readily observable from the northern hemisphere, but unfortunately, if our estimate of the age of its putative nova explosion is correct, it occurred prehistorically.}

However, we suggest that the morphology of the \Ha\ nebula in SY~Cnc argues against the second scenario. The nebula shows no obvious signs of the required deceleration though a collision with the ISM, as it remains very roughly spherical. Additionally, material ejected during an ancient eruption should now lie in a shell with an empty interior, as in fact is seen in AT~Cnc \citep[][their Figure~1]{Shara2012} and V1315~Aql \citep[][their Figures 1--3]{Sahman2018}, or in the form of filaments well separated from the star, as in the Z~Cam nebula \citep[][their Figures 1 and 2]{Shara2007}. Moreover, nebulae ejected by known CN explosions frequently have a bipolar, rather than spherical, morphology; see, for example, the galleries of images of old novae in \citet{Slavin1995} and \citet{Santamaria2020}. But in the case of SY~Cnc, its \Ha\ nebula is approximately round.

3.~Lastly, we briefly outline a possible third scenario for V341~Ara and SY~Cnc, which to our knowledge has not been considered previously. We may be witnessing nova-like CVs that have chanced to encounter interstellar clouds, and while embedded in the clouds the stars underwent CN outbursts. These eruptions ``flash-ionized'' the surrounding gas, producing \Ha\ nebulae centered on the locations of the novae {\it when they occurred}, and which are now recombining. Meanwhile, the CV has moved on, and is now located near the edge of the nebula, where its wind collides with the interstellar gas, producing the bow shocks. The main difference between this scenario and the first one above is the source of ionization for the \Ha\ nebula, either ongoing UV radiation from the CV, or due to the flash of a putative historical nova eruption. In this picture, BZ~Cam has not undergone a CN explosion in the recent past, because no extended off-center \Ha\ nebula is seen. 

In their in-depth study of V341~Ara, CS21 dubbed it the nova-like variable ``that has it all''---bow shocks, nova shells, disk winds, and tilted disks. The main purpose of the present paper is to point out to the CV community a second object that also appears to have it all, and we hope it inspires similar studies in depth.

\acknowledgments


H.E.B. thanks Sakib Rasool for useful discussions and for his lively interest in amateur-astronomer imaging of faint nebulae. 

Preston Starr, Director of the Dark Sky Observatory, has been wonderfully helpful in our work.

The Southern H-Alpha Sky Survey Atlas (SHASSA), was supported by the National Science Foundation.

This work has made use of data from the European Space Agency (ESA) mission
{\it Gaia\/} (\url{https://www.cosmos.esa.int/gaia}), processed by the {\it Gaia\/}
Data Processing and Analysis Consortium (DPAC,
\url{https://www.cosmos.esa.int/web/gaia/dpac/consortium}). Funding for the DPAC
has been provided by national institutions, in particular the institutions
participating in the {\it Gaia\/} Multilateral Agreement.

The Digitized Sky Surveys were produced at the Space Telescope Science Institute under U.S. Government grant NAG W-2166. The images of these surveys are based on photographic data obtained using the Oschin Schmidt Telescope on Palomar Mountain and the UK Schmidt Telescope. The plates were processed into the present compressed digital form with the permission of these institutions. 

We acknowledge with thanks the variable star observations from the AAVSO International Database contributed by observers worldwide and used in this research.

Funding for the \TESS\/ mission is provided by NASA's Science Mission directorate.

This research has made use of the SIMBAD and Vizier databases, operated at CDS, Strasbourg, France.

\bibliography{PNNisurvey_refs}

\end{document}